\documentclass[aps,prb,twocolumn,showpacs,superscriptaddress]{revtex4}
\usepackage{epsfig}
\usepackage{graphicx}
\usepackage{amsfonts}
\usepackage[figuresright]{rotating} 
\usepackage{amssymb}
\usepackage{amsmath}

\newfam\mibfam

  \font\tenmib=cmmib10
  \font\eightmib=cmmib10 scaled 800
  \font\sixmib=cmmib10 scaled 667
  \def\mib{\fam\mibfam\tenmib}
  \textfont\mibfam=\tenmib
  \scriptfont\mibfam=\eightmib
  \scriptscriptfont\mibfam=\sixmib
  
\def\ssr#1{{\sss{\rm #1}}}
\def\sss#1{{\scriptscriptstyle #1}}
\def\ZZ{{\mathbb Z}}
\def\avg#1{\langle#1\rangle}

\def\be{\begin{equation}} \def\ee{\end{equation}}
\def\bea{\begin{eqnarray}} \def\eea{\end{eqnarray}}

\def\nn{\nonumber}

\def\nd{^{\vphantom{\dagger}}}
\def\ns{^{\vphantom{*}}}
\def\yd{^\dagger}
\def\frac#1#2{{\textstyle{#1 \over #2}}}
\def\ket#1{{\big|#1\big\rangle}}
\def\expect#1#2#3{{\big\langle#1\big| #2 \big|#3\big\rangle}}
\def\sket#1{{|#1\rangle}}
\def\bfd{{\mib d}}
\def\bfa{{\mib a}}
\def\bfb{{\mib b}}

\def\bfk{{\mib k}}
\def\bfq{{\mib q}}
\def\bfG{{\mib G}}

\def\bfR{{\mib R}}
\def\ehat{{\hat{\mib e}}}
\def\cP{{\cal P}}
\def\cC{{\cal C}}
\def\cH{{\cal H}}
\def\ie{{\it i.e.\/}}
\def\etc{{\it etc.\/}} 
\def\JH{J_{\rm H}}
\def\JV{J_{\rm V}}
\def\JD{J_{\rm D}}
\def\PsiG{\Psi\nd_{\rm G}}
\def\half{\frac{1}{2}}

\begin{document}

\title{A $\Gamma$-matrix generalization of the Kitaev model}
\author{Congjun Wu}
\affiliation{Department of Physics, University of California, San Diego,
CA 92093}
\author{Daniel Arovas}
\affiliation{Department of Physics, University of California, San Diego,
CA 92093}
\author{Hsiang-Hsuan Hung}
\affiliation{Department of Physics, University of California, San Diego,
CA 92093}
\begin{abstract}
We extend the Kitaev model defined for the Pauli-matrices to the Clifford 
algebra of $\Gamma$-matrices, taking the $4 \times 4$ representation
as an example.  On a decorated square lattice, the ground state spontaneously 
breaks time-reversal symmetry and exhibits a topological phase transition.
The topologically non-trivial phase carries gapless chiral edge modes
along the sample boundary.  On the 3D diamond lattice, the ground states can exhibit gapless 3D
Dirac cone-like excitations and gapped topological insulating states.
Generalizations to even higher rank $\Gamma$-matrices are also discussed.
\end{abstract}
\pacs{75.10.Jm, 73.43.-f, 75.50Mm}
\maketitle

\section{Introduction}
Topological phases of condensed matter have attracted increasing attention in recent years.
Examples include spin liquids, fractional quantum Hall (FQH) states,
and topological insulators, which exhibit fractionalized excitations
and statistics \cite{thouless1982, arovas1984, kohmoto1985, fradkin1986, 
haldane1988, kane2005, kitaev2006, yao2007, feng2007, qixl2008, qixl2008a,
bernevig2006, bernevig2006a}.    A particularly novel application has been the
proposal of utilizing non-Abelian elementary excitations of topological phases
to achieve fault-tolerant topological quantum computation
\cite{kitaev2006,freedman2004}.  To this end, Kitaev, in a seminal paper \cite{kitaev2006}, 
introduced a model of interacting spins on a honeycomb lattice which 
reduces to the problem of Majorana fermions coupled to a static ${\mathbb Z}_2$ gauge field.
The ground state is topologically nontrivial and breaks time reversal (TR) symmetry, and the elementary excitations are anyonic,
with non-Abelian statistics.  Much progress has been made toward understanding the nature of this
phase \cite{yao2007,yu2007,yu2008,yu2008a,lee2007,feng2007,baskaran2007,chen2007,
schmidt2008, dusuel2008, dusuel2008a}, and in generalizing the model to other lattices
\cite{yao2007,si2007,yang2007} and to three dimensions \cite{mandal2008}.
Very recently, a generalization to a decorated honeycomb lattice exhibiting a chiral spin liquid ground state 
was made by Yao {\it et al} \cite{yao2007}.  On the experimental side, the $\nu=\frac{5}{2}$ FQH state is 
expected to realize non-Abelian anyons \cite{willett1987,pan1999,eisenstein2002}.
Time-reversal invariant topological states have also been realized in HgTe semiconductor quantum wells \cite{konig2007} and in Bi$_{1-x}$Sb$_x$ alloys \cite{hsieh2008}.

The solvability of the Kitaev model depends crucially on the property of the Pauli-matrices, {\it i.e.}, 
$\{\sigma_i, \sigma_j\}=2\,\delta_{ij}$ and $\sigma_x \sigma_y \sigma_z=i$,
which are the simplest example of a Clifford algebra.  This gives rise to the constraint that all the above models are defined in lattices with the coordination number three, and thus most of them are
on lattices of dimension two.  Extending the Kitaev model to more general lattices, three
dimensions, and large spin systems enriches this class of solvable topological models.
These extensions naturally involve higher ranked Clifford algebras, with $2^n \times 2^n (n\ge 2)$-dimensional matrices, which can be interpreted as high spin multipole operators.
Some early work on exactly solvable models in the $\Gamma$-matrix representation
of the Clifford algebra has been done in refs. [\onlinecite{wen2003,wen2003a,levin2003}].

In this article, we generalize the Kitaev model from the Pauli matrices
to the Clifford algebra of $\Gamma$-matrices.
For the $4\times 4$ representation, we construct a model
in a decorated square lattice with the coordination number $5$,
which can be interpreted as a spin-$\frac{3}{2}$ magnetic model with
anisotropic interactions involving spin-quadrapole operators only.
It is interesting that although each spin-quadrapole operator 
is TR invariant, the ground state spontaneously breaks TR symmetry.
Such a state is a topologically nontrivial chiral spin liquid state 
with extremely short ranged spin correlation functions.
The topological excitations are expected to be non-Abelian.
The $\Gamma$-matrix formalism is also convenient to define a 3D
counterpart of the Kitaev model on the diamond lattice.
By breaking the TR symmetry explicitly, a gapless spin liquid with
a 3D Dirac-cone like spectrum is found.   Topological insulating states with TR symmetry
also may be elicited on the diamond lattice.
A generalization to even higher rank $\Gamma$-matrices is also
discussed, wherein a topologically non-trivial spin-liquid state 
appears along which manifests a suitable defined ``time-reversal-like'' symmetry.

\section{2D chiral spin liquid with $\Gamma$-matrices}
\subsection{Remarks on the nature of Kitaev's model}
Before elucidating the details of our model, it will prove useful 
to reflect on why the Kitaev model
is equivalent to noninteracting fermions in a static ${\mathbb Z}_2$ 
gauge field.
The $\Gamma$-matrices obeying the Clifford algebra 
$\{\Gamma^a,\Gamma^b\}=2\delta^{ab}$
may be represented in terms of $2n$ Majorana fermions $\eta$ 
and $\{\xi^a\}$, where $a=1,\ldots,2n\!-\!1$.  
Then define the $2n\!-\!1$ $\Gamma$-matrices $\Gamma^a=i\eta\xi^a$.
The product
\be
\Lambda=\Gamma^1\Gamma^2\cdots \Gamma^{2n-1}=i\eta\xi^1\xi^2\cdots\xi^{2n-1}
\label{leqn}
\ee
then commutes with each
of the $\Gamma^a$ and furthermore satisfies $\Lambda^2=(-1)^{n-1}$, 
hence we can choose $\Lambda=\lambda= ( i)^{(n-1)}$, which is to say that all states in each local Hilbert space satisfy
$\Lambda|\Psi\rangle= \lambda |\Psi\rangle $.  Now consider a lattice ${\cal L}$ of coordination number $z=2n\!-\!1$ in which
the link lattice is itself $z$-partite.  That is to say that each link can be assigned one of $z$ colors,
and no two links of the same color terminate in a common site.   On each link $\langle ij\rangle$ of the lattice, then,
we can write interaction terms $\Gamma^a_i \,\Gamma^a_j=-u\nd_{ij}\cdot i\eta\nd_i\eta\nd_j$,
where $i$ and $j$ are the termini of the link, and $u\nd_{ij}=i\xi^a_i\xi^a_j$, where
$a$ is the color label of the link.  Note that $u^2_{ij}=1$, and furthermore 
$[u\nd_{ij},u\nd_{kl}]=0$ for all $\langle ij\rangle$ and $\langle kl\rangle$, and in particular even if
these links share a common terminus.  Thus, the set $\{u\nd_{ij}\}$ thus
defines a configuration for a {\it classical\/}  ${\mathbb Z}_2$ gauge field.   Note that $u\nd_{ij}=-u\nd_{ji}$.
The model, is then
\begin{align}
\cH&=-\sum_{\langle ij\rangle} J\nd_{ij}\,\Gamma^a_i\,\Gamma^a_j\ ,\label{gham}\\
&=\sum_{\langle ij\rangle}J\nd_{ij}\,u\nd_{ij}\cdot i\eta\nd_i\eta\nd_j\ ,\label{mham}
\end{align}
where the `color' index $a$ is associated with each particular link.
The honeycomb lattice satisfies the above list of desiderata, with $z=3$.  Accordingly, the Kitaev model 
has interactions $\sigma^x_i\sigma^x_j$ on $0^\circ$ links, $\sigma^y_i\sigma^y_j$ on $120^\circ$ links, 
and $\sigma^z_i\sigma^z_j$ on $240^\circ$ links.

For $n=2$, the $\Gamma$ matrices are of course the Pauli matrices, and the set  $\{1,\sigma^x,\sigma^y,\sigma^z\}$
forms a basis for rank-2 Hermitian matrices.  For $n=3$, in addition to the five $\Gamma^a$ matrices,
we can define ${2n-1\choose 2}=10$ additional matrices,
\begin{equation}
\Gamma^{ab}\equiv -\frac{i}{2}\,\big[\Gamma^a\,,\,\Gamma^b\big]=-i\xi^a\xi^b\ ,
\label{gamab}
\end{equation}
resulting in a total of $1+5+10=16$, which of course forms a basis for rank-4 Hermitian matrices.  For $n=4$, we
add the ${2n-1\choose 3}=36$ matrices $\Gamma^{abc}\equiv\eta\xi^a\xi^b\xi^c$ to the $\Gamma^{ab}$ (21 total), 
$\Gamma^a$ (7 total), and $1$, resulting in the 64 element basis of rank-8 Hermitian matrices, {\it etc.\/}

For each closed loop $\cC$ on the lattice, one can define a $\ZZ_2$ flux, which is a product,
\be
F\nd_{\cal C}=\prod_{\langle ij\rangle \in{\cal C}} \!\!u\nd_{ij}\ ,
\label{fprod}
\ee
where the product is taken counterclockwise over all links in $\cC$.  If $\cC$ is a self-avoiding loop of $N$ sites (and hence $N$ links), then
\begin{equation}
F\nd_{i\nd_1 i\nd_2\cdots i\nd_N}=-\Gamma^{a\nd_N a\nd_1}_{i\nd_1}\,\Gamma^{a\nd_1 a\nd_2}_{i\nd_2}\cdots
\Gamma^{a\nd_{N-1}a\nd_N}_{i\nd_N}\ .
\end{equation}
These fluxes are all conserved in that they commute with each other and with the Hamiltonian.
Under a local gauge transformation, the Majoranas transform as $\eta\nd_i\to -\eta\nd_i$, which is equivalent to
taking $u\nd_{ij}\to -u\nd_{ij}$ for each link emanating from site $i$.  The gauge-invariant content of the theory thus consists
of the couplings $\{J\nd_{ij}\}$ and the fluxes $\{F\nd_{\cal P}\}$ associated with the elementary plaquettes $\cP$.
For a given set of fluxes, there are many (gauge-equivalent) choices for the $u\nd_{ij}$.   By choosing a particular such gauge
configuration, the Hamiltonian of eqn. \ref{mham} may be diagonalized in a particular sector specified by the $\ZZ_2$ fluxes.

\begin{figure}
\centering\epsfig{file=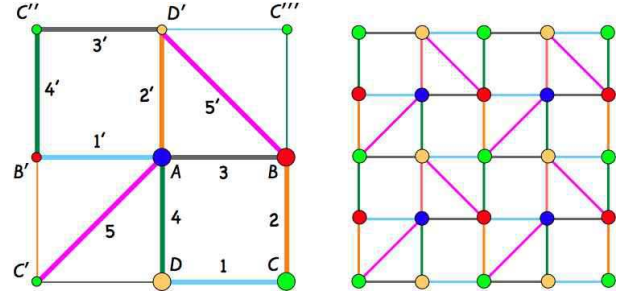,width=8cm}
\caption{(color online) The decorated square lattice (right) with linked diagonal bonds for
the Hamiltonian of eqn. \ref{ham}.  Each unit cell (left) consists four sites (A, B, C, D), ten links, and six plaquettes including 
two squares and four triangles.  The bonds are classified into five types as marked with labels $a$ which run from $1$ to $5$ and $1'$ to $5'$.}
\label{fig:lattice}
\end{figure}

It is worth emphasizing the following features of the $F\nd_\cC$ as defined in eqn. \ref{fprod}.
First, a retraced link contributes a factor of $-1$ to the flux, because $u\nd_{ij}\cdot u\nd_{ji}=-1$.   This has consequences for combining
paths.  If two loops $\cC$ and $\cC'$ share $k$ links in common, then $F\nd_{\cC\cC'}=(-1)^k F\nd_\cC F\nd_{\cC'}$,
where $\cC\cC'$ is the concatenation of $\cC$ and $\cC'$, with the shared links removed.
Consider, for example, the triangles $ABD^\prime$ and 
$BC^{\prime\prime\prime}D^\prime$ in the left panel of fig. \ref{fig:lattice}.  The combination of these triangles yields
the square $ABC^{\prime\prime\prime}D^\prime$.  One then has
\begin{eqnarray}
F\nd_{ABD^\prime}\cdot F\nd_{BC^{\prime\prime\prime}D^\prime}&
=&u\nd_{AB} u\nd_{BD^\prime} u\nd_{D^\prime A}
\cdot u\nd_{BC^{\prime\prime\prime}} u\nd_{C^{\prime\prime\prime}D} u\nd_{DB}\nn \\
&=&u\nd_{AB} u\nd_{BC^{\prime\prime\prime}} u\nd_{C^{\prime\prime\prime}D^\prime} 
u\nd_{D^\prime A} \cdot(-1)\nn \\
 &=&-F\nd_{ABC^{\prime\prime\prime}D^\prime}\ ,
\end{eqnarray}
since the single link $BD$ is held in common, but is traversed in opposite directions.

A second point regarding the fluxes $F\nd_\cC$ is that if $n(\cC)$ is the number of links contained in $\cC$, then traversing $\cC$
clockwise rather than counterclockwise results in a flux of $F\nd_{\cC^{-1}}=(-1)^{n(\cC)} F\nd_\cC$.  Thus, the $\ZZ_2$ flux reverses
sign for odd length loops if the loop is traversed in the opposite direction.  For loops of even length, the flux is invariant under the direction
of traversal.

\subsection{Model definition}
Our generalization of Kitaev's model involves an $n=3$ system on the decorated square lattice depicted in fig. \ref{fig:lattice}.
The link lattice is $5$-partite: each lattice node lies at the confluence of $5$ differently colored links.  We will take the side length of each square to be $a$.  The lattice constant for the underlying Bravais lattice, \ie\ the distance between two closest A sublattice
sites, is then $a'=2a$.  The model is that of eqn. \ref{gham}:
\begin{align}
\cH&=-J_1 \sum_{\rm 1\,links} \Gamma^1_i \,\Gamma^1_j
-J_2 \sum_{\rm 2\,links} \Gamma^2_i\, \Gamma^2_j   \\
&\quad -J_3 \sum_{\rm 3\,links} \Gamma^3_i \,\Gamma^3_j 
-J_4 \sum_{\rm 4\,links} \Gamma^4_i \,\Gamma^4_j \nn -J_5 \sum_{\rm 5\,links} 
\Gamma^5_i \,\Gamma^5_j. 
\label{ham}
\end{align}
While the structural unit cell shown in the left panel of fig. \ref{fig:lattice} contains distinct bonds, labeled with $a$ among
$\{1,\ldots,5\}$ or $\{1',\ldots,5'\}$, and is solvable with 10 distinct couplings $J\nd_a$, we shall assume $J\nd_1=J\nd_{1'}$ {\it etc.\/}.  

The $4\times 4$ $\Gamma$-matrices may be explicitly taken as
\be
\Gamma^1=i\begin{pmatrix} 0 & -{\mathbb I} \\ {\mathbb I} & 0 \end{pmatrix}
\ ,\ \Gamma^{2,3,4}=\begin{pmatrix} {\vec\sigma} & 0 \\ 0 & -{\vec\sigma} \end{pmatrix}
\ ,\  \Gamma^5=\begin{pmatrix} 0 & {\mathbb I} \\ {\mathbb I} & 0 \end{pmatrix} \ ,
\label{4x4gammas}
\ee
where ${\mathbb I}$ and $\vec{\sigma}$ are the $2\times 2$ unit and 
Pauli matrices, respectively.  These five $\Gamma$-matrices are in fact spin-quadruple 
operators for the spin-$\frac{3}{2}$ system \cite{murakami2004}.
The other ten $\Gamma$-matrices, defined above in eqn. \ref{gamab},
contains both spin and spin octupole operators \cite{wu2006a,wu2003}.
For later convience, the ten $\Gamma^{ab}$ matrices are also written
explicitly here as
\begin{align}
&\Gamma^{12,13,14}&=\begin{pmatrix} 0 & {\vec\sigma} \\ {\vec\sigma} & 0 \end{pmatrix} \quad  &,\ &
\Gamma^{25,35,45}&=i\begin{pmatrix} 0 & -{\vec\sigma} \\ {\vec\sigma} & 0 \end{pmatrix}&\label{4x4gammas_2}\\
&\Gamma^{34,42,23}&=\begin{pmatrix} {\vec\sigma} & 0 \\ 0 & {\vec\sigma} \end{pmatrix} \quad  &,\ &
\Gamma^{51}&=\begin{pmatrix} {\mathbb I} & 0 \\ 0 & -{\mathbb I} \end{pmatrix}\ .&\nonumber
\end{align}

From the six Majorana fermions, we may fashion three Dirac fermions, {\it viz.}
\bea
c\nd_{04}&=&\frac{1}{2} (\eta + i\xi^4) \\
c\nd_{15}&=&\frac{1}{2} (\xi^1 - i\xi^5) \\
c\nd_{23}&=&\frac{1}{2} (\xi^2 - i \xi^3)\ .
\eea
The three diagonal $\Gamma$-matrices can be represented as
\bea
\Gamma^4&=&  2 c\yd_{04} c\nd_{04}-1\\
\Gamma^{15}&=& 2 c\yd_{15} c\nd_{15}-1\\
\Gamma^{23}&=& 2 c\yd_{23} c\nd_{23}-1\ .
\eea
The basis vectors of the physical space in terms of the $S^z$-eigenstates are:
\begin{align}
\ket{+\frac{3}{2}}&=c\yd_{23} c^\dagger_{15} c^\dagger_{04}\,\ket{\Omega} \\
\ket{+\frac{1}{2}}&=i c\yd_{15}\,\ket{\Omega}\\
\ket{-\frac{1}{2}}&=c\yd_{23} \,\ket{\Omega}\\
\ket{-\frac{3}{2}}&=-i c\yd_{04} \,\ket{\Omega}\ .
\end{align}
where $\sket{\Omega}$ is the reference vacuum state.
In other words, the physical states have odd fermion occupation number
$n=c\yd_{04} c\nd_{04} + c\yd_{15} c\nd_{15} +c\yd_{23} c\nd_{23}$.
The $R$-matrix defined in Sect. \ref{subsect:tr} can be represented as
$R=-i (c\nd_{15} + c\yd_{15}) (c\nd_{23} -c\yd_{23})$.
 
For each of the six plaquettes per unit cell, one defines a $\ZZ_2$ flux as in eqn. \ref{fprod}.
For example, for the $ABD^\prime$ triangle shown in fig. \ref{fig:lattice} 
enclosed by the $a=2$, $a=3$, and $a=5$ bonds, we have
\begin{align}
F\nd_{\!ABD^\prime}&=\big(i\xi^3_A\xi^3_B\big)\big(i\xi^5_B\xi^5_{D^\prime}\big)
\big(i\xi^2_{D^\prime}\xi^2_A\big)\nn\\
&=-\Gamma^{23}_A\,\Gamma^{35}_B\,\Gamma^{52}_{D^\prime}\ .
\end{align}
These fluxes are all conserved in that they commute with each other and with the Hamiltonian.
As pointed in refs. [\onlinecite{yao2007,kitaev2006}], the flux configuration for the ground
state on the triangular lattice is odd under TR symmetry, thus the ground state breaks TR symmetry and
is at least doubly degenerate.

\subsection{Time reversal}
\label{subsect:tr}
The time reversal (TR) transformation is defined as the product  $T=RC$,
where $C$ is the complex conjugation operator and $R$ is the charge conjugation
matrix, satisfying $R^2=-1$ and $R\yd=R^{-1}=R^{\rm t}=-R$.  Explicitly, we can take
$R=\Gamma^1\Gamma^3=\xi^1\xi^3$.  Note that $R\,\Gamma^a\, R=-(\Gamma^a)^{\rm t}$
and $R\,\Gamma^{ab}\, R= (\Gamma^{ab})^t$.  Acting on the Majoranas, the
complex conjugation operator $C$ is defined so that
\be
C\left\{
\begin{array}{c}
\eta\\ \xi^{1,3}\\ \xi^{2,4,5}
\end{array}
\right\}C
=\left\{
\begin{array}{c}
\eta\\ \xi^{1,3}\\ -\xi^{2,4,5}
\end{array}
\right\}
\ee
With these definitions, the $\Gamma^a$ operators are even under TR while the $\Gamma^{ab}$ operators are odd.  Note that $R\eta R=\eta$.  Thus, the effect
of the time reversal operation on the fluxes $F\nd_\cC$ is the same as that of
reversing the direction in which the loop is traversed, \ie\ $R F\nd_\cC R
= (-1)^{n(\cC)}F\nd_\cC$, where $n(\cC)$ is the number of links in $\cC$.

\subsection{Projection onto the physical sector}
As we have seen, in terms of the Majorana fermions, the $\Gamma$-matrices
are represented as
\bea
\Gamma^a=i \eta \xi^a\quad,\quad \Gamma^{ab}=-i \xi^a \xi^b.
\eea
We further demand that any physical state $|\Psi\rangle$ must satisfy $\Lambda\nd_i|\Psi\rangle = -|\Psi\rangle$,
where $\Lambda\nd_i=\Gamma_i^1\,\Gamma_i^2\,\Gamma_i^3\,\Gamma_i^4\,\Gamma_i^5$, on each
lattice site $i$ (see eqn. \ref{leqn}).  That is, each state in the
eigenspectrum is also an eigenstate of the projector $P=\prod_i \frac{1}{2}(1-\Lambda_i)$,
so the local Hilbert space at each site is four-dimensional, rather than eight-dimensional.
Since $[H,P]=0$, the two operators can be simultaneously diagonalized.  For any eigenstate
$\sket{\Psi}$ of $H$ in the extended Hilbert space (with local dimension eight), we have that
$P\sket{\Psi}$ is also an eigenstate of $H$, and with the same eigenvalue.  Thus, we are free
to solve the problem in the extended Hilbert space, paying no heed to the local constraints,
and subsequently apply the projector $P$ to each eigenstate of $H$ if we desire the actual
wavefunctions or correlation functions.  Note also that while the projector $P$ {\it does not commute\/}
with the link $\ZZ_2$ gauge fields $u\nd_{ij}$, nevertheless $[P,F\nd_{\cal C}]=0$ for any closed loop ${\cal C}$,
so the projector {\it does\/} commute with all the $\ZZ_2$ fluxes, which, aside from the couplings, 
constitute the gauge-invariant content of the Hamiltonian.

\subsection{General Majorana Hamiltonian}
The general noninteracting lattice Majorana Hamiltonian is written as $\cH=\half\sum_{i,j} H\nd_{ij}\,\eta\nd_i\,\eta\nd_j$,
with $H=H\yd= -H^*$ and $\{\eta\nd_i,\eta\nd_j\}=2\delta\nd_{ij}$.
Let $\bfR$ denote a Bravais lattice site, \ie\ a site on the $A$ sublattice, and the index
$l\in\{1,\ldots,2r\}$ labels a basis element; our model has $r=2$. The Majorana fermions satisfy
$\{\eta\nd_l(\bfR),\eta\nd_{l'}(\bfR')\}=2\,\delta\nd_{\bfR\bfR'}\,\delta\nd_{ll'}$, and
Fourier transforming to $\eta\nd_l(\bfk)=N^{-1/2}\sum_\bfR\eta\nd_l(\bfR)\,
e^{-i\bfk\cdot(\bfR+\bfd\ns_l)}$, where $N$ is the number of unit cells and
$\bfd\ns_l$ is the location in the unit cell of the $l^{\rm th}$ basis site,
we arrive at the Hamiltonian 
\begin{equation}
\cH=\half\sum_\bfk H\nd_{ll'}(\bfk)\,\eta\nd_l(-\bfk)\,\eta\nd_{l'}(\bfk)\ ,
\end{equation}
where $H\nd_{ll'}(\bfk)=\sum_\bfR H\nd_{ll'}(\bfR-\bfR')\,e^{i\bfk\cdot(\bfR'-\bfR+\bfd\ns_{l'}-\bfd\ns_{l})}$
satisfies
\begin{equation}
H\nd_{ll'}(\bfk)=H^*_{l'l}(\bfk)=-H\nd_{l'l}(-\bfk)=-H^*_{ll'}(-\bfk)\ ,
\label{hconds}
\end{equation}
and $\eta\nd_l(-\bfk)=\eta\yd_l(\bfk)$.  The eigenvalues of $H(\bfk)$ occur in pairs
$\big\{E\nd_j(\bfk)\,,\,-E\nd_j(-\bfk)\big\}$,
where $j\in\{1,\ldots,r\}$ where $2r$ is the number of basis elements.  Written in terms of Dirac fermions,
the diagonalized Hamiltonian takes the form
\begin{equation}
\cH=\sum_\bfk\sum_{j=1}^r \big(2\gamma\yd_{j,\bfk}\,\gamma\nd_{j,\bfk}-1\big) E\nd_j(\bfk)\ ,
\end{equation}
and therefore the ground state energy is
\begin{equation}
E\nd_0=-\sum_\bfk\sum_{j=1}^r \big| E\nd_j(\bfk) \big|
\end{equation}
and the excitation energies are $\omega\nd_j(\bfk)=2\big| E\nd_j(\bfk) \big|$.

\subsection{Bulk and edge spectra}
Following the procedure outlined by Kitaev\cite{kitaev2006},
we represent the Hamiltonian $H$ in the extended Hilbert space in terms of free Majorana
fermions hopping in the presence of a static $\ZZ_2$ gauge field, as in eqn. \ref{mham}.
The spectra can be solved in each gauge sector with a specified 
distribution of the $\ZZ_2$ phases $u_{ij}=\pm 1$.

In the extended Hilbert space, all the link phases $u\nd_{ij}$ mutually commute with each other and
with the Hamiltonian.  The product of the link phases around a given plaquette gives the $\ZZ_2$ flux
associated with that plaquette, as in eqn. \ref{fprod}.  It is the $\ZZ_2$ fluxes of all the triangular and square
plaquettes which define the gauge-invariant content of the model.

\begin{figure}
\centering\epsfig{file=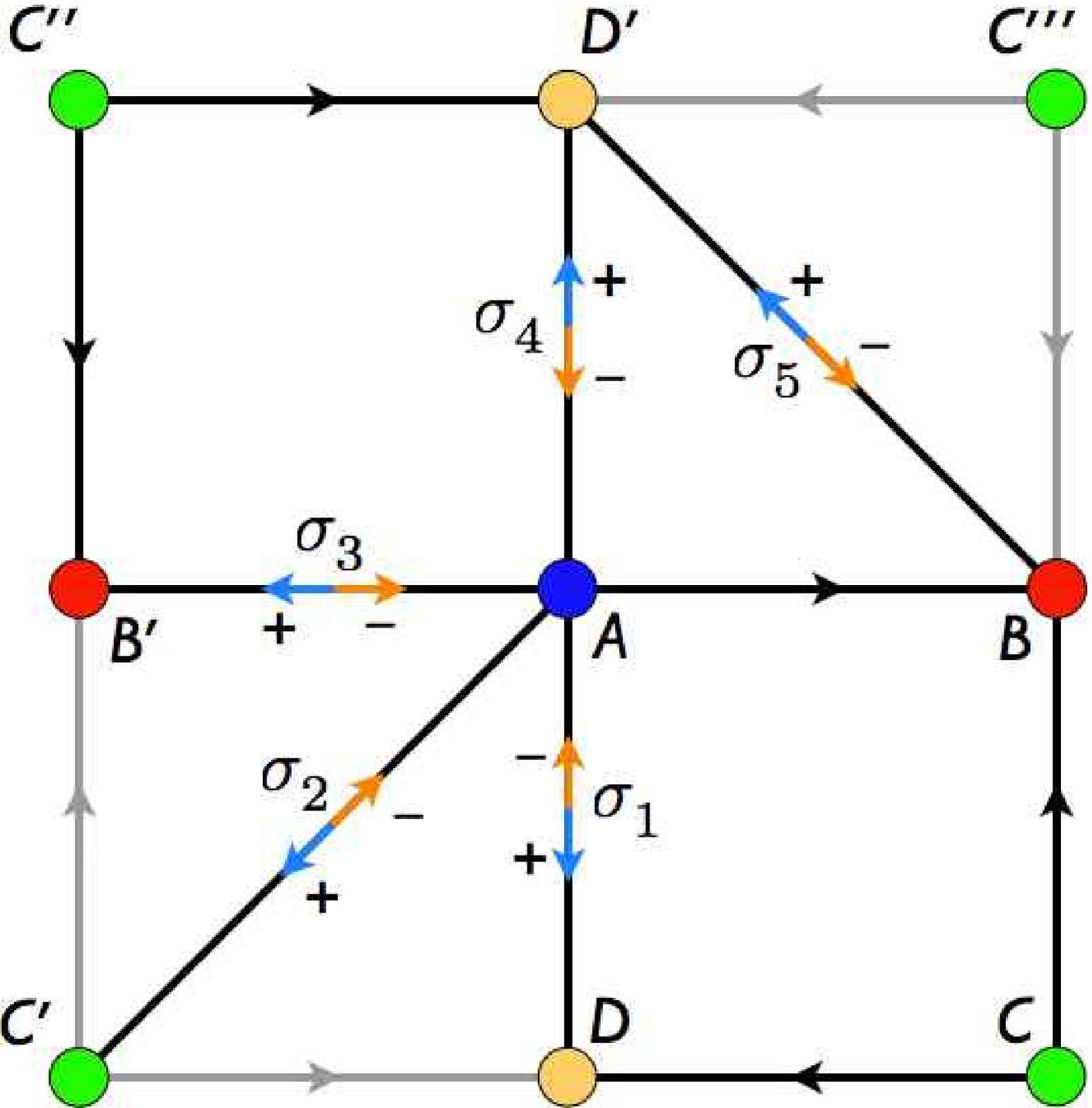,width=6cm}
\caption{(color online) When the magnetic and structural unit cells coincide, there
five $\ZZ_2$ gauge degrees of freedom, which we take to be $\sigma_{1-5}$ as shown, with
$u_{AD}=\sigma_1$, \etc\  The other $u_{ij}$ are positive for links where the black arrow points from $i$ to $j$.}
\label{fig:gauge}
\end{figure}

The {\it structural\/} unit cell, as shown in the left panel of fig. \ref{fig:lattice}, consists of four sites, ten links, and
six plaquettes.  Suppose that the arrangement of fluxes $F\nd_{\cal C}$ has the same period as this structural cell.
What of the link phases $u\nd_{ij}$?  It is easy to see that if the total $\ZZ_2$ flux of the structural unit cell is $+1$,
{\it i.e.\/} if the product of the $F\nd_{\cal C}$ over the six plaquettes in the structural unit cell is $+1$,
then the $u\nd_{ij}$ may be chosen so as to be periodic in this unit cell.   In such a case the {\it magnetic unit cell\/}
coincides with the structural unit cell.\cite{footnote1}  If, however, the net $\ZZ_2$ flux per structural unit cell is
$-1$, then the smallest magnetic unit cell ({\it i.e.\/} periodic arrangement of the link phases $u\nd_{ij}$) necessarily
comprises two structural cells.  Here, we assume that the net flux per structural cell is $+1$, so the magnetic and
structural unit cells coincide.  Thus, there are five (and not six) $\ZZ_2$ degrees of freedom per unit cell, which,
using the labels of fig. \ref{fig:gauge}, can be taken to be the link phases
\begin{align}
u\nd_{AD}\equiv\sigma\nd_1 \quad,\quad u\nd_{AC'}&=\sigma\nd_2 \quad,\quad u\nd_{AB'}=\sigma\nd_3\\
u\nd_{AD'}=\sigma\nd_4\quad &,\quad u\nd_{BD'}=\sigma\nd_5\ .
\end{align}
The remaining five values of $u\nd_{ij}$ can then be fixed, and we take
\begin{equation}
u\nd_{AB}=u\nd_{CB}=u\nd_{C''B'}=u\nd_{CD}=u\nd_{C''D'}=1\ .
\end{equation}
The fluxes of the triangular and square plaquettes are then given by
\begin{align*}
F\nd_{ABCD}&=\sigma\nd_1 & F\nd_{BC'''D'}&=\sigma\nd_5\\
F\nd_{AD'C''B'}&=\sigma\nd_3\,\sigma\nd_4 & F\nd_{C'DA}&=-\sigma\nd_1\,\sigma\nd_2\\
F\nd_{ABD'}&=-\sigma\nd_4\,\sigma\nd_5 &F\nd_{AB'C'}&=\sigma\nd_2\,\sigma\nd_3\ .
\end{align*}
Thus, there are $2^5=32$ possible distinct flux configurations which are periodic in this unit cell.
Any arrangement of the $u_{ij}$ phases consistent with zero total flux per cell is therefore
identical or gauge-equivalent to one of these 32 configurations.

We set the coupling on the horizontal links to $\JH$, on the vertical links to $\JV$, and
on the diagonal links to $\JD$.  The independent nonzero elements of $H\nd_{ll'}(\bfk)$ are then
\begin{align}
H\nd_{AB}(\bfk)&=i\,\JH\big(e^{i\theta\ns_1/2}+\sigma\ns_3\,e^{-i\theta\ns_1/2}\big)\\
H\nd_{AC}(\bfk)&=i\,\JD\,\sigma\ns_2\,e^{-i(\theta\ns_1+\theta\ns_2)/2}\\
H\nd_{AD}(\bfk)&=i\,\JV\big(\sigma\ns_4\,e^{i\theta\ns_2/2} + \sigma\ns_1\,e^{-i\theta\ns_2/2}\big)\\
H\nd_{BC}(\bfk)&=-2i\,\JV\,\cos(\theta\ns_2/2)\\
H\nd_{BD}(\bfk)&=i\,\JV\,\sigma\ns_5\,e^{i(\theta\ns_1-\theta\ns_2)/2}\\
H\nd_{CD}(\bfk)&=2i\,\JH\,\cos(\theta\ns_1/2)\ ,
\end{align}
where we define the angles $(\theta\nd_1,\theta\nd_2)=(k\nd_x a',k\nd_y a')$.

We have found that the ground state energy is minimized when the $\ZZ_2$ flux through each of the four squares
in the unit cell is $F\nd_\Box=-1$, and the triangles all are the same value, \ie\ $F\nd_\bigtriangleup=\pm 1$.
Under time reversal $F\nd_\Box\to F\nd_\Box$ and $F\nd_\bigtriangleup\to -F\nd_\bigtriangleup$.  Note that
the $\ZZ_2$ flux through either square which contains a diagonal bond is $F\nd_\Box=-F^2_\bigtriangleup=-1$
if the triangles contain the same flux.  The flux pattern with $F\nd_\Box=F\nd_\bigtriangleup=-1$ is achieved 
with the choice
\begin{equation}
(\sigma\nd_1,\sigma\nd_2,\sigma\nd_3,\sigma\nd_4,\sigma\nd_5)=(-1,-1,+1,-1,-1)\ .
\end{equation}
With this link flux assignment, the Hamiltonian matrix $H(\bfk)$ takes the form
\begin{widetext}
\begin{equation}
H(\bfk)=\begin{pmatrix}
0 & 2i\,\JH\,\cos(\theta\nd_1/2) & -i\,\JD\,e^{-i(\theta\nd_1+\theta\nd_2)/2} & -2i\,\JV\,\cos(\theta\nd_2/2) \\
-2i\,\JH\,\cos(\theta\nd_1/2) & 0 & -2i\,\JV\,\cos(\theta\nd_2/2) & -i\,\JD\, e^{i(\theta\nd_2-\theta\nd_1)/2} \\ 
i\,\JD\,e^{i(\theta\nd_1+\theta\nd_2)/2} & 2i\,\JV\,\cos(\theta\nd_2/2) & 0 & 2i\,\JH\,\cos(\theta\nd_1/2) \\ 
2i\,\JV\,\cos(\theta\nd_2/2) & i\,\JD\, e^{i(\theta\nd_1-\theta\nd_2)/2}  & -2i\,\JH\,\cos(\theta\nd_1/2) & 0
\end{pmatrix}\ .\label{eq:band}
\end{equation}
\end{widetext}

\begin{figure}[!b]
\centering\epsfig{file=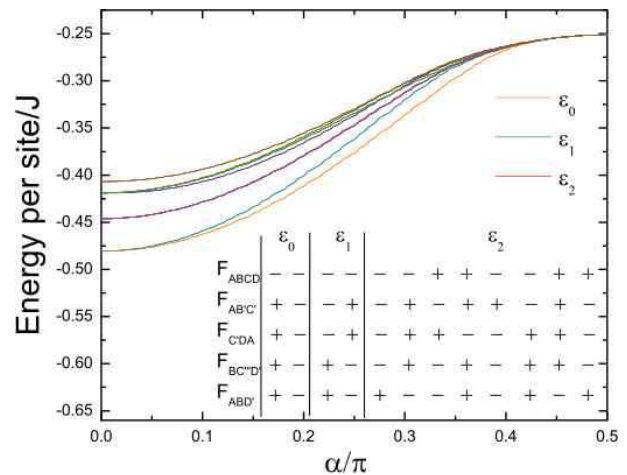,width=8cm}
\caption{(color online) Total energies (per site) {\it versus\/} $\alpha=\tan^{-1}(J_\ssr{D}/J)$ for our decorated square lattice model
Curves for all possible flux configurations are shown.  The flux configurations for the two lowest-lying total energy states are twofold
degenerate owing to time-reversal, which reverses the flux in the odd-membered loops.  The third-lowest lying total energy state
has an eightfold degenerate flux configuration.  At $\alpha=\half\pi$ the horizontal and vertical bond strengths vanish and the
system becomes a set of disconnected dimers.}
\label{fig:total}
\end{figure}

\begin{figure}[!t]
\centering\epsfig{file=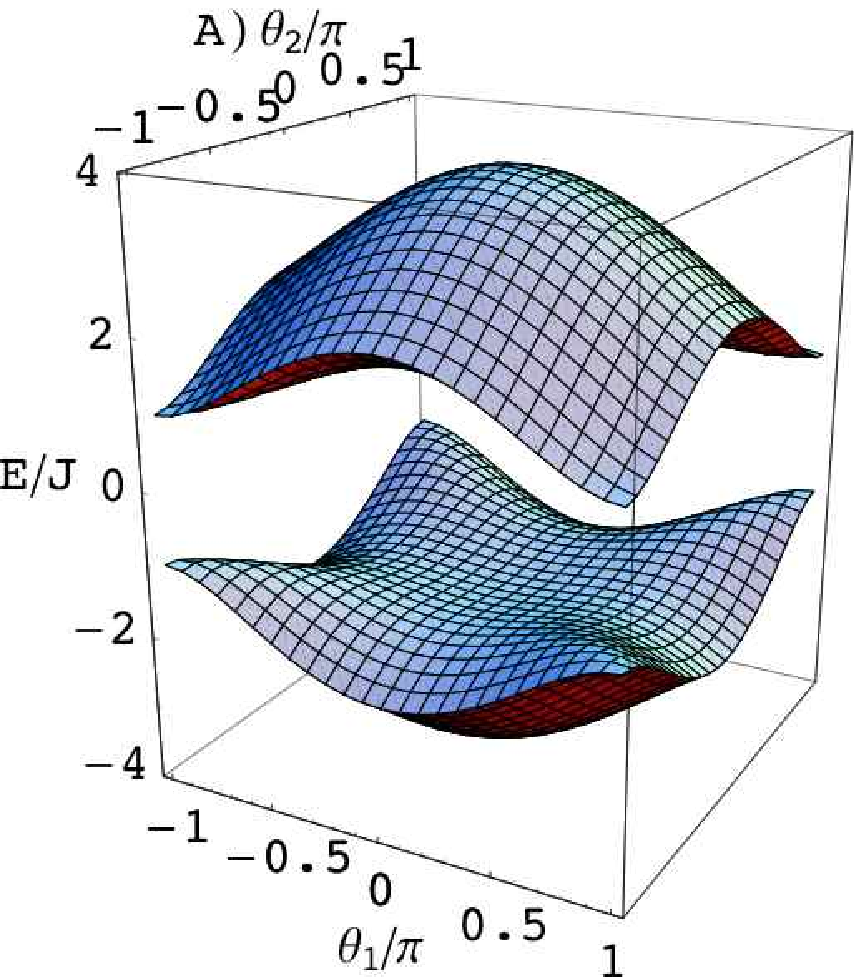,clip=1,width=0.48\linewidth,angle=0}
\centering\epsfig{file=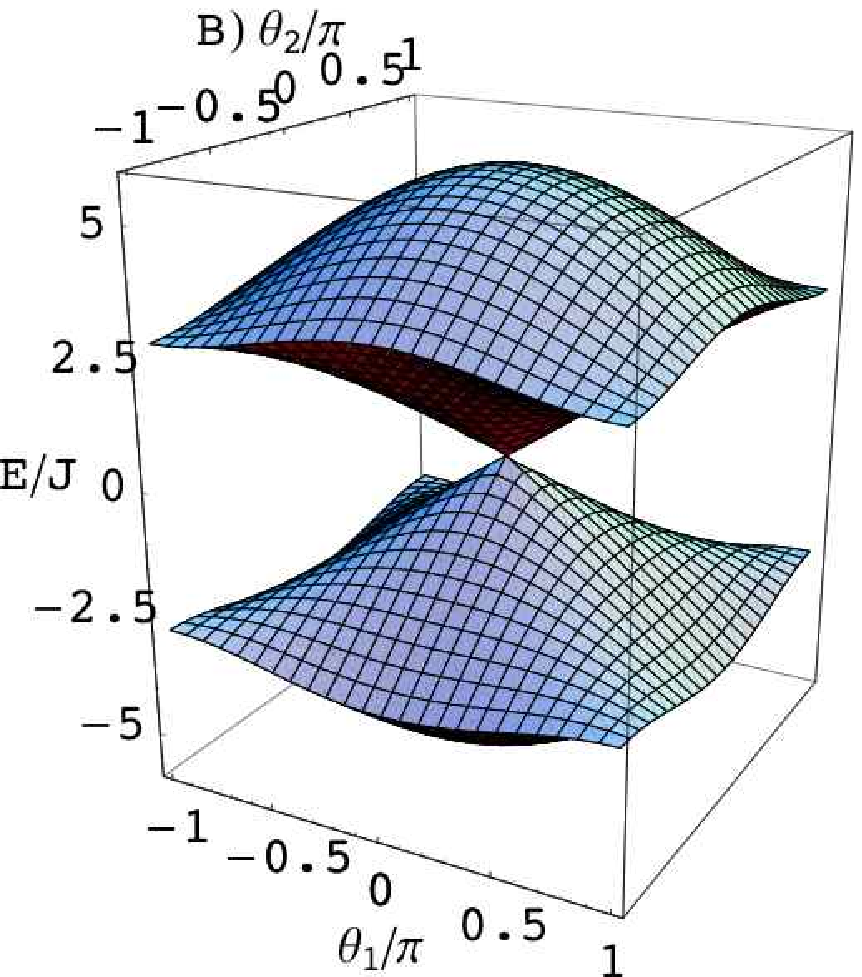,clip=1,width=0.48\linewidth,angle=0}
\centering\epsfig{file=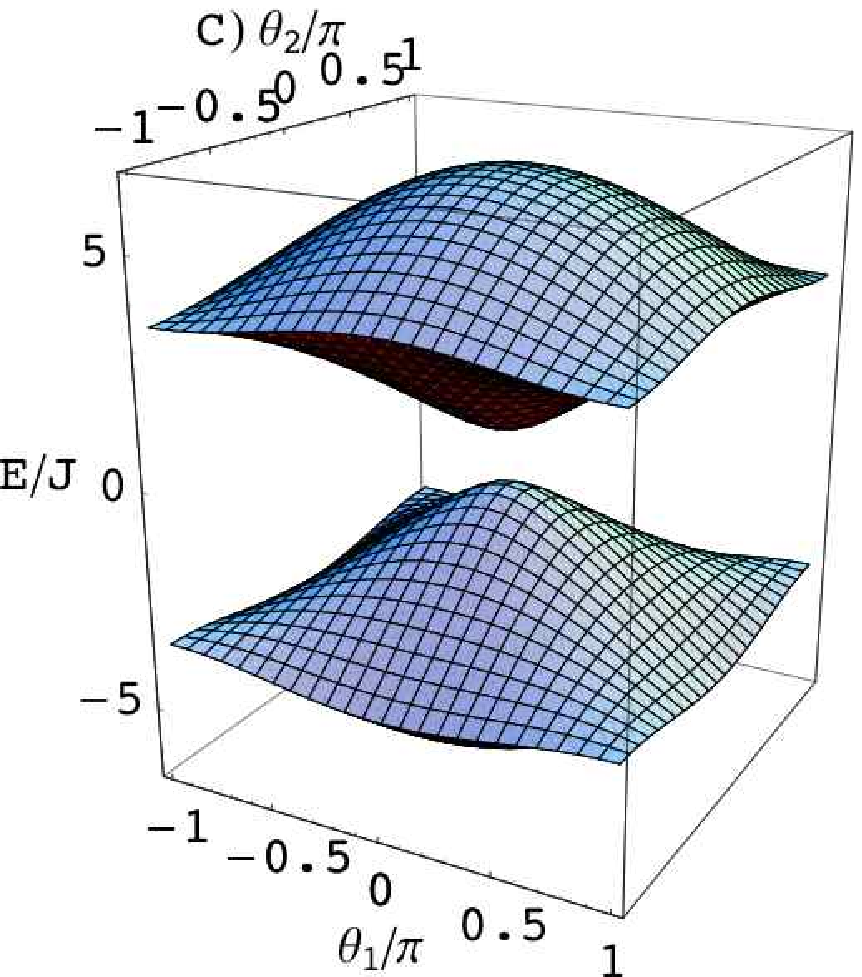,clip=1,width=0.48\linewidth,angle=0}
\caption{The band structure of Eq. \ref{eq:band} at $\JD/J>0$.
(A) $\JD/J=1$ with a massive Dirac spectrum at $(\pi,\pi)$
with the gap value $2 \JD$;  (B) $\JD/J=2\sqrt 2$ where a gapless Dirac
cone appears; (C) $\JD/J =3.5$.   The gap value at $(0,0)$ is $2|\JD-2\sqrt 2J|$.
A topological phase transition occurs from a topological nontrivial
phase at $\JD/J<2\sqrt 2$ to a trivial phase at $\JD/J>2\sqrt 2$.}
\label{fig:energy}
\end{figure}

The eigenvalues of $H(\bfk)$ are found to be
\begin{align}
E\ns_{1,2}(\bfk)&=-2\sqrt{\frac{J_{\rm D}^2}{4} + (\JH^2 + \JV^2) f(\bfk)  \pm \JD\sqrt{\JH^2+\JV^2}\,g(\bfk)}\\
E\ns_{3,4}(\bfk)&=+2\sqrt{\frac{J_{\rm D}^2}{4} + (\JH^2 + \JV^2) f(\bfk)  \mp \JD\sqrt{\JH^2+\JV^2}\,g(\bfk)}    \ ,
\end{align}
where
\begin{align}
f(\bfk)&={\JH^2\cos^2\!\big(\half\theta\ns_1\big)+\JV^2\cos^2\!\big(\half\theta\ns_2\big)\over
\JH^2+\JV^2}\\
g(\bfk)&=\cos\big(\half\theta\ns_1\big)\cos\big(\half\theta\ns_2\big)\ .
\end{align}
In fig. \ref{fig:total} we plot the total energy per site for our model for all possible flux configurations for our model
(32 in total), where we have taken $\JH=\JV=J$, which we here and henceforth assume.  We explore the properties
of our model as a function of the dimensionless parameter $\JD/J$.   Since time-reversal has the effect of sending $F\to -F$ 
or all odd-membered loops, every state must have an even-fold degeneracy.  We find that the flux configurations with
the two lowest lying total energies are each twofold degenerate, and the third-lowest lying total energy flux configuration
is eightfold degenerate.  In fig. \ref{fig:energy} we show the spectra $E\ns_j(\bfk)$ for the lowest energy flux configuration.

\begin{figure}
\centering\epsfig{file=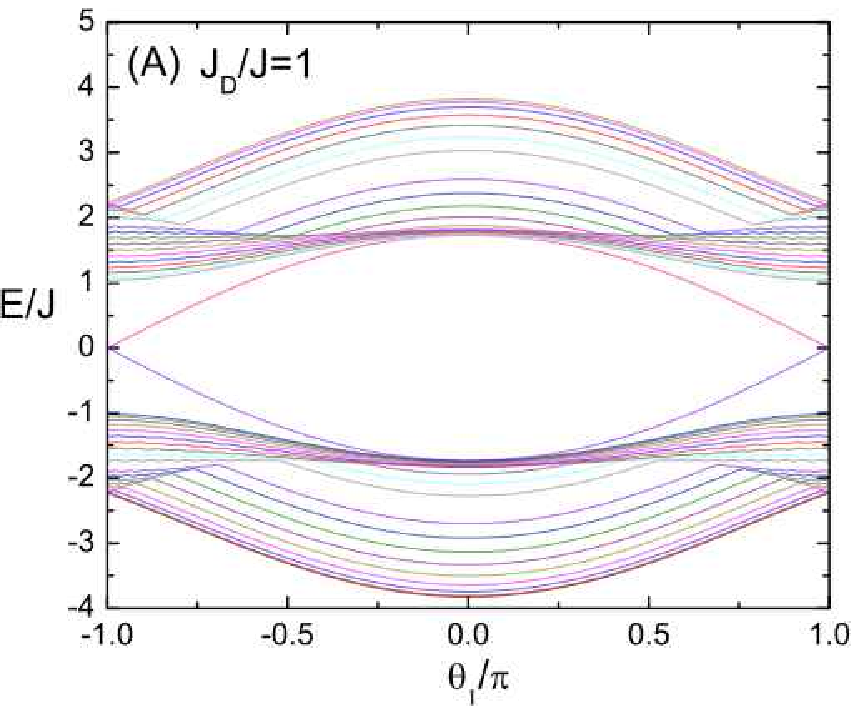,clip=1,width=0.9\linewidth,angle=0}
\centering\epsfig{file=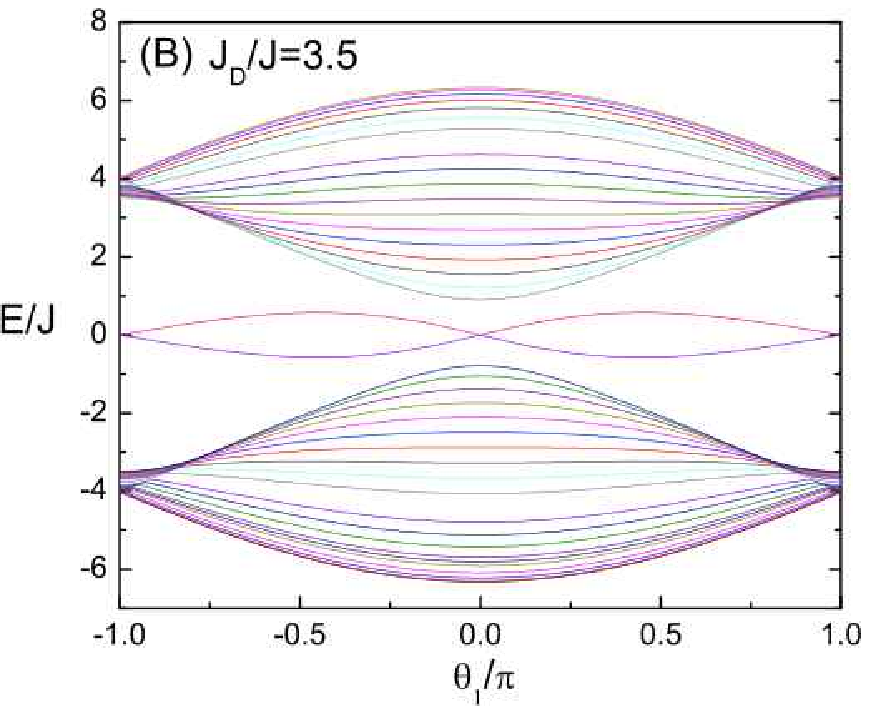,clip=1,width=0.9\linewidth,angle=0}
\caption{A) The gapless edge modes of $\eta$-fermions across the band gap appear in open boundary samples
in the topologically nontrivial phase at $\JD/J<2\sqrt 2$.  B) The edge modes become nontopological at $\JD/J>2\sqrt 2$.}
\label{fig:edge}
\end{figure}

Our model exhibits a topological phase transition as increasing $\JD/J$. 
At $\JD/J=0$, the dispersion is gapless, with a Dirac cone at $(\theta\nd_1,\theta\nd_2)=(\pi,\pi)$.
As $\JD/J \neq 0$, this spectrum becomes massive with a gap $\Delta(\pi,\pi) = 2 \JD$.
This is a topologically nontrivial phase, characterized by non-vanishing
Chern numbers, with concomitant gapless edge modes inside the gap between
bands 2 and 3 in a sample with an open boundary, as depicted in Fig. \ref{fig:edge}(A).
As $\JD/J$ increases, the second and third bands are pushed
toward each other at the Brillouin zone center.  These bands eventually touch when
$\JD/J=2 \sqrt 2$, forming there a new gapless Dirac cone.
For $\JD/J >2\sqrt 2$, the system is in a topologically trivial phase 
in which the edge states no longer exhibit a spectral flow, and they
remain confined within the band gap, as shown in Fig. \ref{fig:edge}(B).
In this case, the edge states are no longer topologically protected, and they are
sensitive to local perturbations, and indeed can be eliminated by sufficiently strong
such perturbations.  The gap located at $(\theta\nd_1,\theta\nd_2)=(0,0)$ is found to be
$\Delta(0,0)= 2 | \JD -2\sqrt 2 J|$.

\subsection{Spin correlation functions}
As is the case with Kitaev's model \cite{baskaran2007}, the spin correlation 
function in our ground state is short-ranged due to the conserved flux of each plaquette.
We take the ground state wavefunction $\ket{\PsiG} = P \ket{\Psi}$ where $\ket{\Psi}$ is the unprojected state,
with each $u^a_{ij}$ taking a value $\pm 1$.  Then $\expect{\PsiG}{\Gamma^a_i \,\Gamma^b_j}{\PsiG}
=\expect{\Psi}{P \,\eta_i \eta_j\, \xi^a_i \xi^b_j}{\Psi}$.
Unless site $i$ and $j$ are linked, and unless $a=b$ appropriate to the bond
$\avg{ij}$, it is always possible to find a loop whose flux is flipped
by $\xi^a_i$ or $\xi^b_j$.  In that case, since $\eta_i \eta_j$ and $P$ do not change the loop flux, 
the above expectation value vanishes.  A similar reasoning shows that any two point correlation 
of the type of $\expect{\PsiG}{\Gamma_i^a\, \Gamma^{cd}_j}{\PsiG}$
or $\expect{\PsiG}{\Gamma_i^{ab}\, \Gamma^{cd}_j}{\PsiG}$ vanishes as well.

Generally speaking, a non-local spin correlation function is finite only if,
in the Majorana representation of its operators, the $\xi$'s
can be expressed as a product of the ${\mathbb Z}_2$ gauge phases defined on different links, {\it i.e.\/}
\bea
\prod_{\langle i j \rangle} \xi^a_i \xi^a_j=\prod_{\langle i j\rangle}(-i u^a_{ij})
\eea
Similarly to the work of ref. \cite{yao2007}, it is natural to expect that
each topological excitation of a $\ZZ_2$-vortex traps an unpaired Majorana mode.
These vortex excitations will exhibit non-Abelian statistics.  We are now performing numerical
calculations to confirm this prediction.

\section{An Extension to 8 $\times$ 8 $\Gamma$-matrices}
The above procedure is readily extended to even higher rank Clifford
algebras, such as the $n=4$ case with eight Majorana fermions and seven 
anticommuting $8\times 8$ $\Gamma$ matrices.  We choose the Majoranas
to be $\{\eta^r,\xi^a\}$ where $r=1,2,3$ and $a=1,2,3,4,5$.  We write
$\Gamma^a=i\eta^1\xi^a$ as before, and we define $\Gamma^{6a}=i\eta^2\xi^a$.
The Hamiltonian is
\begin{align}
\cH&=-\sum_{a=1}^5 J\nd_a\sum_{a\ {\rm links}}
\big(\Gamma^a_i\, \Gamma^a_j-\Gamma^{6a}_i\, \Gamma^{6a}_j\big)\\
&=\sum_{\langle ij\rangle}J\nd_{ij}\,u\nd_{ij}\cdot i\,(\eta^1_i\eta^1_j-\eta^2_i\eta^2_j)\ .
\end{align}
Physical states are projected, at each site, onto eigenstates of the operator
$\Lambda=i\eta^1\eta^2\eta^3\xi^1\xi^2\xi^3 \xi^4\xi^5$
with eigenvalue $+i$.

We obtain the same behavior of $\eta^{1,2}$ modes as before,
and a flat band of zero energy mode of $\eta^{3}$.
Within a fixed gauge choice of $u_{ij}$,  $\cH$ is invariant 
under a suitably defined TR reversal-like transformation of 
as ${\tilde T} \eta_1{\tilde T}^{-1}= \eta_2$, ${\tilde T}\eta_2 {\tilde T}^{-1}= -\eta_1$,
which is not related to the physical TR transformation.
In the topologically non-trivial phase, the $\eta^{1,2}$ edge modes 
have opposite chirality, which is robust in the absence of a
perturbation of the type $\Gamma^{6}\equiv i\eta^1\eta^2$, which
breaks the symmetry under ${\tilde T}$.
This behavior is similar to that of the ${}^3$He-B phase in two dimensions,
which has been recently identified as a TR-invariant topological superconductor \cite{qixl2008}.

\section{$\Gamma$-matrix models in the 3D diamond lattice}
\begin{figure}
\centering\epsfig{file=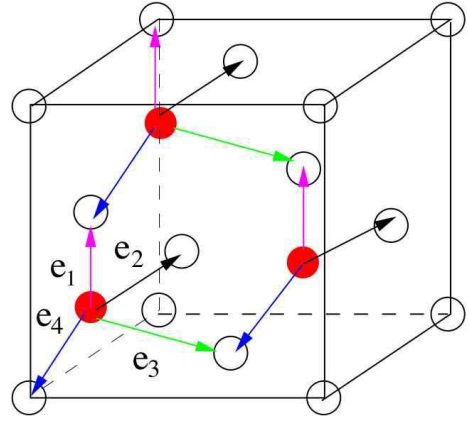,clip=1,width=0.7\linewidth,angle=0}
\caption{The 3D diamond lattice with the nearest separation
$a$, and two sublattices $A$ (filled circles) and $B$ (hollow circles).
The unit vectors $\ehat_{1,2,3,4}$ are defined from each A site
to its four B neighbors (see text).}
\label{fig:diamond}
\end{figure}

The $\Gamma$-matrix analogy of the Kitaev model is also extendable 
to the three-dimensional (3D) systems.  We will consider here a diamond lattice, which is fourfold coordinated,
as illustrated in Fig. \ref{fig:diamond}.  
In the following, we will first describe a model with a 3D Dirac
cone and broken TR symmetry, and then another one exhibiting
3D topological insulating states with TR symmetry maintained.

\subsection{3D Spin liquid with Dirac cone excitations}
We first consider an $n=3$ model ($4\times 4$ $\Gamma$-matrices)  in the 
diamond lattice with explicity time reversal symmetry breaking.
Recall that the diamond lattice is bipartite, consisting of two FCC sublattices.
Each $A$ sublattice site is located in the center of a tetrahedron of $B$ sites, and {\it vice versa\/},
as shown in fig. \ref{fig:diamond}.  We define the unit vectors
\begin{align}
\ehat\nd_1&=\frac{1}{\sqrt{3}}(-1,1,1) & \ehat\nd_3&=\frac{1}{\sqrt{3}}(1,1,-1) \\
\ehat\nd_2&=\frac{1}{\sqrt{3}}(1,-1,1) & \ehat\nd_4&=\frac{1}{\sqrt{3}}(-1,-1,-1)\ ,\nonumber
\end{align}
which point from a given $A$ sublattice site to its four $B$ sublattice neighbors.  The basis vectors for
the underlying FCC Bravais lattice, which can be taken to be the $A$ sublattice itself, are then
\begin{align}
\bfa\nd_1&=\ehat\nd_1-\ehat\nd_4=\frac{2}{\sqrt{3}}(0,1,1)\nonumber\\
\bfa\nd_2&=\ehat\nd_2-\ehat\nd_4=\frac{2}{\sqrt{3}}(1,0,1)\\
\bfa\nd_3&=\ehat\nd_3-\ehat\nd_4=\frac{2}{\sqrt{3}}(1,1,0)\nonumber \ .
\end{align}
A general $A$ sublattice site lies at $\bfR=n\nd_1\bfa\nd_1 + n\nd_2\bfa\nd_2 + n\nd_3\bfa\nd_3$
The two-element diamond unit cell may be taken to consist of the $A$ site at $\bfR$ and the $B$ site at
$\bfR+\ehat\ns_4$.  It is also useful to define the null vector $
\bfa\nd_4\equiv \ehat\nd_4-\ehat\nd_4=0$.

The Brillouin zone of the fcc-Bravais lattice is a dodecahedron.
The elementary reciprocal lattice vectors, which form a basis for a BCC lattice, are
\begin{align}
\bfb\nd_1&=\frac{\sqrt{3}}{2}\pi\,(-1,1,1)\nonumber\\
\bfb\nd_2&=\frac{\sqrt{3}}{2}\pi\,(1,-1,1)\\
\bfb\nd_3&=\frac{\sqrt{3}}{2}\pi\,(1,1,-1)\nonumber\ ,
\end{align}
and satisfy $\bfa\nd_i\cdot\bfb\nd_j=2\pi\delta\nd_{ij}$, where $i,j\in \{1,2,3\}$.
Any vector in $\bfk$-space may be decomposed in to components in this basis, {\it viz.\/}
\begin{equation}
\bfk={\theta\ns_1\over 2\pi}\,\bfb\ns_1 + {\theta\ns_2\over 2\pi}\,\bfb\ns_2 + {\theta\ns_3\over 2\pi}\,\bfb\ns_3\ .
\end{equation}
We then have $\bfk\cdot\bfa\nd_a=\theta\ns_a$, with $\theta\ns_4=0$.

We begin with the Heisenberg-like Hamiltonian,
\begin{align}
\cH\nd_0&=-\sum_{a=1}^4 J\ns_a\sum_\bfR \Gamma^a_\bfR\,{\tilde\Gamma}^a_{\bfR+\ehat\ns_a}\\
&=\sum_\bfR\sum_{a=1}^4 J\ns_a\, u^a_\bfR\cdot i\, \eta\nd_\bfR\,{\tilde\eta}\nd_{\bfR+\ehat\ns_a}\ .
\end{align}
We use the tilde to distinguish operators which reside on the $B$ sublattice from those which reside
on the $A$ sublattice.  Here we have defined
\begin{equation}
u^a_\bfR=i\,\xi^a_\bfR\,{\tilde\xi}^a_{\bfR+\ehat\ns_a}
\end{equation}
as the $\ZZ_2$ gauge field on the link between $\bfR$ and $\bfR+\ehat\ns_a$.

\begin{figure}
\centering\epsfig{file=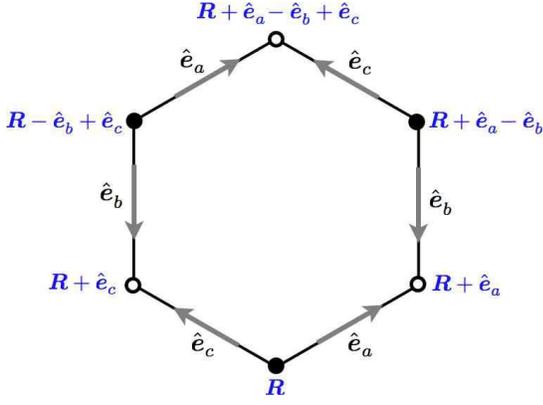,width=2.8in}
\caption{A non-coplanar hexagonal ring within the diamond lattice.}
\label{fig:hexface}
\end{figure}

Within the diamond lattice, one can identify four classes of hexagonal loops.  Starting at any $A$ sublattice site, we can
move in a cyclical six-site path by traversing consecutively the displacement vectors
$\{\ehat\ns_a\,,\,-\ehat\ns_b\,,\,\ehat\ns_c\,,\,-\ehat\ns_a\,,\,\ehat\ns_b\,,\,-\ehat\ns_c\}$.
We label this hexagon by the indices $\langle abc\rangle$.  Without loss of generality we can assume $a<b<c$,
since any permutation of these indices results in the same loop, traversed in either the same or the opposite sense,
depending on whether the permutation is even or odd, respectively.  
According to Lieb's theorem \cite{lieb1994}, the gauge flux  in the ground state is $-1$ in each such loop, thus we can set
$u^a_\bfR=1$ on each link, for each $a\in\{1,2,3,4\}$, because a circuit around a six-site loop involves three $AB$ links,on which the $\ZZ_2$ gauge field is $u\ns_{ij}=+1$, and three $BA$ links, on which $u\ns_{ij}=-1$.
Consider now
\begin{equation}
\Gamma^a_\bfR\, {\tilde\Gamma}^{ab}_{\bfR+\ehat\ns_a} \Gamma^b_{\bfR+\ehat\ns_a-\ehat\ns_b}=
-u^a_\bfR\,u^b_{\bfR+\ehat\ns_a-\ehat\ns_b}\!\cdot i\eta\ns_\bfR\,\eta\ns_{\bfR+\ehat\ns_a-\ehat\ns_b}
\end{equation}
and
\begin{equation}
{\tilde\Gamma}^a_{\bfR+\ehat\ns_a} \Gamma^{ab}_{\bfR}\> {\tilde\Gamma}^b_{\bfR+\ehat\ns_b}=
-u^a_\bfR\,u^b_\bfR\cdot i{\tilde\eta}\ns_{\bfR+\ehat\ns_a}\,{\tilde\eta}\ns_{\bfR+\ehat\ns_b}\ .
\end{equation}
These operators both break time-reversal symmetry, owing to the presence of the $\Gamma^{ab}$ and
${\tilde\Gamma}^{ab}$ factors.  However, owing to the commuting nature of the $\ZZ_2$ link variables $u^a_\bfR$,
we may add terms such as these to $\cH\ns_0$ and still preserve the key feature that the Hamiltonian describes
Majorana fermions $\eta$ and ${\tilde\eta}$ hopping in the presence of a {\it static\/} $\ZZ_2$ gauge field.

To recover the point group symmetry of the underlying lattice, we sum over contributions on each loop
$\langle abc\rangle$.  We define the operators
\begin{align}
V_\bfR^{ab}&=\Gamma^a_\bfR\, {\tilde\Gamma}^{ab}_{\bfR+\ehat\ns_a} \Gamma^b_{\bfR+\ehat\ns_a-\ehat\ns_b}\\
{\tilde V}_\bfR^{ab}&={\tilde\Gamma}^a_{\bfR+\ehat\ns_a} \Gamma^{ab}_{\bfR}\> {\tilde\Gamma}^b_{\bfR+\ehat\ns_b}\ .
\end{align}
The time-reversal violating term in our Hamiltonian is then written as
\begin{equation}
\cH\nd_1=\sum_\bfR\sum_{a<b}^4 \Big( h\ns_{ab}\,V_\bfR^{ab} + {\tilde h}\ns_{ab}\,{\tilde V}_\bfR^{ab} \Big)\ .
\end{equation}
We reiterate that this model is also exactly solvable due to the commutativity of the link fluxes $u^a_\bfR$,
and we assume $u^a_\bfR=1$ for all $a$ and $\bfR$.

Transforming to $\bfk$ space, we have the Hamiltonian matrix
\begin{equation}
H\ns_{ll'}(\bfk)=\begin{pmatrix} \omega(\bfk) & \Delta(\bfk) \\
\Delta^*(\bfk) & -{\tilde\omega(\bfk)} \end{pmatrix}
\end{equation}
where, after performing a unitary transformation to remove a phase $\half e^{\mp i(\theta\ns_1+\theta\ns_2+\theta\ns_3)/4}$
from the diagonal terms, we have
\begin{align}
\omega(\bfk)&=\sum_{a<b}^4 h\ns_{ab}\,\sin(\theta\ns_a-\theta\ns_b)\\
{\tilde\omega}(\bfk)&=\sum_{a<b}^4 {\tilde h}\ns_{ab}\,\sin(\theta\ns_a-\theta\ns_b)\\
\Delta(\bfk)&=i\sum_{a=1}^4 J\ns_a\,e^{i\theta\ns_a}\ .
\end{align}

\begin{figure}
\centering\epsfig{file=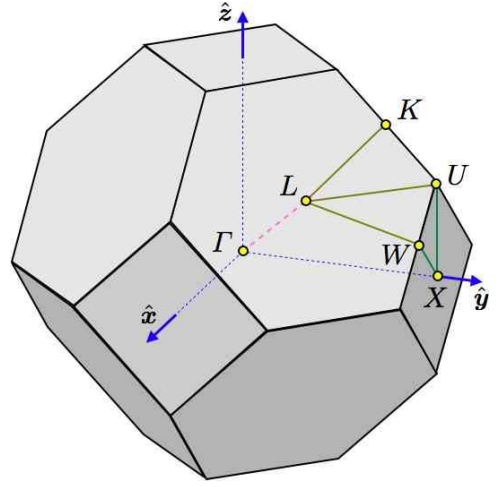,width=2.5in}
\caption{First brillouin zone for the FCC structure.}
\label{fig:fccbzone}
\end{figure}

The eigenvalues are $E\ns_\pm(\bfk)=\pm\sqrt{ \omega^2(\bfk)+ |\Delta(\bfk)|^2}$.  One finds that $E\ns_\pm(\bfk)$ both vanish when
$\bfk$ lies at one of the three inequivalent $X$ points, which lie at the centers of the square faces of the first Brillouin zone,
at locations $\half(\bfb\ns_2+\bfb\ns_3)$, $\half(\bfb\ns_1+\bfb\ns_3)$, and $\half(\bfb\ns_1+\bfb\ns_2)$. 
Expanding about the last of these, we write $\theta\ns_1=\pi+\psi\ns_1$, $\theta\ns_2=\pi+\psi\ns_2$, and
$\theta\ns_3=\psi\ns_3$, and assuming $J\ns_a=J$ and ${\tilde h}\ns_{ab}=h\ns_{ab}$, we find to lowest order in
$\psi\ns_{1,2,3}$ that
\begin{align} 
\omega(\bfk)&=(h^{12}-h^{13}-h^{14})\,\psi\ns_1 -  (h^{12}-h^{23}+h^{24})\,\psi\ns_2\nonumber \\
&\qquad\qquad +  (h^{13}+h^{23}+h^{34})\,\psi\ns_3\\
\Delta(\bfk)&=J\,(\psi\ns_1+\psi\ns_2-\psi\ns_3)\ .\nonumber
\end{align}
If we write $\bfk=\half(\bfb\ns_1+\bfb\ns_2)+\bfq$, then we have
\begin{align}
q\ns_x&=\frac{\sqrt{3}}{4}\,\big(\!-\psi\ns_1 + \psi\ns_2 + \psi\ns_3\big) \\
q\ns_y&=\frac{\sqrt{3}}{4}\,\big(\psi\ns_1 - \psi\ns_2 + \psi\ns_3\big) \\
q\ns_z&=\frac{\sqrt{3}}{4}\,\big(\psi\ns_1 + \psi\ns_2 - \psi\ns_3\big) \ .
\end{align}
Thus, the term $\omega(\bfk)={\tilde\omega}(\bfk)$ can be written as $h\,\bfk\cdot{\hat{\mib g}}\ns_\perp$,
where ${\hat{\mib g}}\ns_\perp$ is a unit vector lying in the $(x,y)$ plane, and $h$ is some combination of the
$h\ns_{ab}$.  The spectrum is therefore that of a deformed Dirac cone, linear in the two directions ${\hat{\mib g}}\ns_\perp$ 
and ${\hat{\mib z}}$, and quadratic in the third.
These deformed Dirac cones can be made gapped by introducing anisotropy
in the $J$-terms, say, $J_4 \ne J_{1,2,3}\equiv J$.
The system may then become a 3D topological insulator with TR
symmetry breaking by developing chiral surface states.

At this point we must ask: whither the $\xi^5$ Majoranas?
Indeed, for the Hamiltonian $\cH=\cH\ns_0+\cH\ns_1$, the $\xi^5$ Majoranas form a flat band at zero energy.
To provide a dispersion for the $\xi^5$ branch, we define a new Hamiltonian $\cH^{\xi^5}$ which we form
from $\cH^\eta$
with the replacements $\Gamma^a \to \Gamma^{5a}$.  Equivalently, we can skip to the final form of $\cH$ in terms
of $\eta\ns_\bfR$ and ${\tilde\eta}\ns_{\bfR+\ehat\ns_4}$ and replace them with $\xi^5_\bfR$ and
${\tilde\xi}^5_{\bfR+\ehat\ns_4}$, respectively.  Combining these individual Hamiltonians to form
$\cH=\cH^\eta-\cH^{\xi^5}$ results in a matrix $H(\bfk)$ of the form
\begin{align}
H(\bfk)&=\begin{pmatrix} \omega(\bfk) & 0 & \Delta(\bfk) & 0 \\
0 & -\omega(\bfk) & 0 & -\Delta(\bfk) \\
\Delta^*(\bfk) & 0 & -\omega(\bfk) & 0 \\
0 & -\Delta^*(\bfk) & 0 & \omega(\bfk) \end{pmatrix} \label{eq:4by4}\\
&=\omega(\bfk)\,\gamma^4 - {\rm Re}\,\Delta(\bfk)\,\gamma^{14} + {\rm Im}\,\Delta(\bfk)\,\gamma^{45}\ ,\nonumber
\end{align}
where the row and column indices range over $\eta$, $\xi^5$, ${\tilde\eta}$, and ${\tilde\xi}^5$, consecutively.
The $\gamma$-matrices here are used as a basis for $4\times 4$ Hermitian 
matrices, which are not to be confused with the spin and spin-multipole
operators denoted as $\Gamma$-matrices.
Nevertheless we still choose $\gamma^a$ and $\gamma^{ab}$ 
taking the same values as the matrix forms of $\Gamma^a$ and $\Gamma^{ab}$
defined in eqns. \ref{4x4gammas} and \ref{4x4gammas_2}.

Again the spectra of Eq. \ref{eq:4by4} exhibits gapless Dirac cones at
the three $X$-points. This gapless excitation is robust if there no mixing terms between
$\eta$ and $\xi^5$ fermions, which would appear as terms involving 
$\{\gamma^2,\gamma^3,\gamma^{24},\gamma^{34}\}$, corresponding to same sublattice hopping,
or $\{\gamma^{12},\gamma^{13},\gamma^{25},\gamma^{35}\}$, corresponding to alternate sublattice hopping.
The introduction of such couplings can produce a gap in the spectrum and give rise to spin liquid states.

\subsection{Time reversal symmetry properties}
The Hamiltonian matrix $H(\bfk)$ is Hermitian, and can therefore be expanded as
\begin{equation}
H(\bfk)=\sum_{a=1}^5 \lambda\ns_a(\bfk)\,\gamma^a + \sum_{a<b}^5 \lambda\ns_{ab}(\bfk)\,\gamma^{ab}\ ,
\end{equation}
where the couplings $\lambda\ns_a$ and $\lambda\ns_{ab}$ are real functions of their arguments.
Due to the conditions of eqn. \ref{hconds}, we see that these couplings come in two classes.
The coefficients of purely real, symmetric $\gamma$-matrices must satisfy
$\lambda(\bfk)=-\lambda(-\bfk)$; we call this class odd, or $-$.
The coefficients of the purely imaginary, antisymmetric $\gamma$-matrices must satisfiy
$\lambda(\bfk)=\lambda(-\bfk)$; we call this class even, or $+$.  Thus, we have
\begin{align}
\hbox{\rm class $-$}\ :\ &{\mathbb I}\,,\,\lambda\ns_2\,,\,\lambda\ns_4\,,\,\lambda\ns_5\,,\,\lambda\ns_{12}\,,\,\lambda\ns_{14}\,,\,\lambda\ns_{23}
\,,\,\lambda\ns_{34}\,,\,\lambda\ns_{15}\,,\,\lambda\ns_{35}\nonumber\\
\hbox{\rm class $+$}\ :\ &\lambda\ns_1\,,\,\lambda\ns_3\,,\,\lambda\ns_{13}\,,\,\lambda\ns_{24}\,,\,\lambda\ns_{25}\,,\,\lambda\ns_{45}\ .
\end{align}
The periodicity under translations $\bfk\to\bfk+\bfG$ through a reciprocal lattice vector
then requires $\lambda(\bfG/2)=\lambda(-\bfG/2)=0$ for the odd class.
For three-dimensional systems, there are eight wavectors within the first Brillouin zone which satisfy this condition,
\ie\ $\bfk=\frac{1}{2}(n_1 \bfb_1+n_2 \bfb_2 +n_3 \bfb_3)$ for $n_i=0$ or $1$, which are identified as
the zone center $\Gamma$, the four inequivalent $L$-points, and the three inequivalent $X$-points (see fig. \ref{fig:fccbzone}).

If we consider $(\eta\,,\,\xi^5)$ as a Kramers doublet of
pseudo-spin up and down, we can define a TR-like anti-unitary
transformation ${\cal T}$ as
\bea
{\cal T}= i \gamma^{24}\, {\cal C} ,
\eea
where ${\cal C}$ is the complex-conjugation operation.  
Under this operation, we have
\begin{equation}
{\cal T} \>\eta\>{\cal T}^{-1} = \xi^5 \qquad,\qquad
{\cal T} \>\xi^5\>{\cal T}^{-1} =-\eta\ .
\end{equation}
The $\gamma$-matrices then divide into two classes
under time reversal: even (${\cal T}\,\gamma\,{\cal T}^{-1}=\gamma$) or odd
(${\cal T}\,\gamma\,{\cal T}^{-1}=-\gamma$).  We find
\begin{align}
\hbox{\rm ${\cal T}$-even}\ :\ & {\mathbb I}\,,\gamma^5,\gamma^{15}, \gamma^{25}, \gamma^{35}, \gamma^{45}\\
\hbox{\rm ${\cal T}$-odd}\ :\ & \gamma^1, \gamma^2, \gamma^3,  \gamma^4,\gamma^{12},
 \gamma^{13}, \gamma^{14}, \gamma^{23},  \gamma^{24},\gamma^{34}\ .\nonumber
\end{align}
Since ${\cal T}$ also reverses the direction of $\bfk$, sending $\bfk\to -\bfk$, time reversal symmetry requires
that the coefficients of the ${\cal T}$-even $\gamma$-matrices satisfy $\lambda(\bfk)=\lambda(-\bfk)$, while
the coefficients of the ${\cal T}$-odd $\gamma$-matrices must satisfy $\lambda(\bfk)=-\lambda(-\bfk)$.
Taking into account the division into even and odd classes, we find that time reversal symmetry,
\ie\ ${\cal T}\cH{\cal T}^{-1}=\cH$, requires the vanishing of the following coefficients:
\begin{align}
\lambda\ns_1=&\,\lambda\ns_3=\lambda\ns_{13}=\lambda\ns_{15}=\lambda\ns_{24}=\lambda\ns_{35}=0\\
&(\hbox{\rm forbidden by ${\cal T}$-symmetry})\ .\nonumber
\end{align}

We may also define a parity operator ${\cal P}$, as
\begin{equation}
{\cal P}=\gamma^{45}\,{\cal I}\ .\,
\end{equation}
where ${\cal I}$ is the lattice inversion operator which inverts the coordinates relative to the position ${\mib r}=\half\ehat\ns_4$.
Under ${\cal P}$ we then have the classification
\begin{align}
\hbox{\rm ${\cal P}$-even}\ :\ & {\mathbb I}\,,\gamma^1, \gamma^2, \gamma^3, \gamma^{12}, \gamma^{13}, \gamma^{23}, \gamma^{45}\\
\hbox{\rm ${\cal P}$-odd}\ :\ & \gamma^4, \gamma^5, \gamma^{14},  \gamma^{15},\gamma^{24},
 \gamma^{25}, \gamma^{34}, \gamma^{35}\ .\nonumber
\end{align}
Parity also acts on crystal momentum, sending $\bfk\to -\bfk$.  As a result, 
\begin{align}
\lambda\ns_2&=\lambda\ns_{12}=\lambda\ns_{24}=\lambda\ns_{25}=\lambda\ns_{34}=0\\
&(\hbox{\rm forbidden by ${\cal P}$-symmetry})\ .\nonumber
\end{align}
Note that the product ${\cal PT}$ does not reverse $\bfk$, and is given by
\begin{equation}
{\cal PT}=\gamma^{25}\,{\cal CI}\ .
\end{equation}
A summary of the symmetry properties of the different $\gamma$-matrices is provided in table \ref{gammatab}.

\begin{table}[!t]
\begin{center}
\begin{tabular}{||c|c|c|c|c||c|c|c|c|c||}
\hline\hline
$\gamma$ & class & ${\cal T}$ & ${\cal P}$ & ${\cal PT}$ & $\gamma$ & class & ${\cal T}$ & ${\cal P}$ & ${\cal PT}$\\ 
\hline\hline
${\mathbb I}$ & $-$ & $+$ & $+$ & $+$ & $\gamma^{14}$ & $-$ & $-$ & $-$ & $+$ \\ \hline
$\gamma^1$ & $+$ & $-$ & $+$ & $-$ & $\gamma^{15}$ & $-$ & $+$ & $-$ & $-$ \\ \hline
$\gamma^2$ & $-$ & $-$ & $+$ & $-$ & $\gamma^{23}$ & $-$ & $-$ & $+$ & $-$ \\ \hline
$\gamma^3$ & $+$ & $-$ & $+$ & $-$ & $\gamma^{24}$ & $+$ & $-$ & $-$ & $+$ \\ \hline
$\gamma^4$ & $-$ & $-$ & $-$ & $+$ & $\gamma^{25}$ & $+$ & $+$ & $-$ & $-$ \\ \hline
$\gamma^5$ & $-$ & $+$ & $-$ & $-$ & $\gamma^{34}$ & $-$ & $-$ & $-$ & $+$ \\ \hline
$\gamma^{12}$ & $-$ & $-$ & $+$ & $-$ & $\gamma^{35}$ & $-$ & $+$ & $-$ & $-$ \\ \hline
$\gamma^{13}$ & $+$ & $-$ & $+$ & $-$ & $\gamma^{45}$ & $+$ & $+$ & $+$ & $+$ \\ \hline
\hline
\end{tabular}
\caption{\label{gammatab} Symmetry properties of the $\gamma$-matrices.}
\end{center}
\end{table}

\subsection{3D topological states with TR symmetry}
Recently, topological insulating states in three dimensions have attracted
a great deal of attention \cite{fu2007, fu2007a, moore2007,
roy2006a, roy2006b, roy2008, ryu2008}.
Below we will show by  adding the hybridization between 
$\eta$ and $\xi^5$ fermions, we can arrive at the gapped
spin liquid states which can be topologically nontrivial.

We begin with the additional hybridization term,
\begin{align}
\Delta\cH&=\sum_\bfR\Bigg\{ m\Big(\Gamma^5_\bfR + {\tilde\Gamma}^5_{\bfR+\ehat\ns_4}\Big) +\\
&\qquad\quad + \sum_{a \ne b}^4 g\ns_{ab}\,\Gamma^a_\bfR\,{\tilde\Gamma}^{ab}_{\bfR+\ehat\ns_a}
\Gamma^{b5}_{\bfR+\ehat\ns_a-\ehat\ns_b}\nonumber\\
&\qquad\qquad\qquad +\sum_{a \ne b}^4 {\tilde g}\ns_{ab}\,{\tilde\Gamma}^a_{\bfR+\ehat\ns_a}\,\Gamma^{ab}_\bfR
\ {\tilde\Gamma}^{b5}_{\bfR+\ehat\ns_b}\Bigg\}\ .\nonumber
\end{align}
Note that
\begin{align}
\Gamma^a_\bfR\,{\tilde\Gamma}^{ab}_{\bfR+\ehat\ns_a} \Gamma^{b5}_{\bfR+\ehat\ns_a-\ehat\ns_b}&=
u^a_\bfR\,u^b_{\bfR+\ehat\ns_a-\ehat\ns_b}\cdot i\eta\ns_\bfR\,\xi^5_{\bfR+\ehat\ns_a-\ehat\ns_b}\\
{\tilde\Gamma}^a_{\bfR+\ehat\ns_a}\,\Gamma^{ab}_\bfR\ {\tilde\Gamma}^{b5}_{\bfR+\ehat\ns_b}&=
u^a_\bfR\,u^b_\bfR\cdot i\eta\ns_{\bfR+\ehat\ns_a}\,\xi^5_{\bfR+\ehat\ns_b}\ .
\end{align}
Assuming once again that $u_\bfR^a=1$ for all $\bfR$ and $a$, and further taking ${\tilde g}\ns_{ab}=g\ns_{ba}$,
this leads to the matrix Hamiltonian
\begin{align}
H(\bfk)&=\begin{pmatrix} \omega(\bfk) & \beta(\bfk) & \Delta(\bfk) & 0 \\
\beta^*(\bfk) & -\omega(\bfk) & 0 & -\Delta(\bfk) \\
\Delta^*(\bfk) & 0 & -\omega(\bfk) &  \beta(\bfk) \\
0 & -\Delta^*(\bfk) & \beta^*(\bfk) & \omega(\bfk) \end{pmatrix} \label{top4by4}\\
&=\omega(\bfk)\,\gamma^4 + {\rm Re}\,\Delta(\bfk)\,\gamma^{14} - {\rm Im}\,\Delta(\bfk)\,\gamma^{45}\nonumber\\
&\qquad\qquad\qquad + {\rm Re}\,\beta(\bfk)\,\gamma^{34} + {\rm Im}\,\beta(\bfk)\,\gamma^{24}\ ,\nonumber
\end{align}
where
\begin{equation}
\beta(\bfk)=im + i\sum_{a,b} g\ns_{ab}\>e^{i(\theta\ns_a-\theta\ns_b)}\ .
\end{equation}
The spectra for each momentum $\bfk$ is doubly degenerate, and the two degenerate energy levels are found to be 
\bea
E\ns_\pm(\bfk)=\pm\sqrt{\omega^2(\bfk) + |\Delta(\bfk)|^2 + |\beta(\bfk)|^2}\ .
\eea
In the presence of the ${\cal T}$ symmetry, ${\rm Im}\,\beta(\bfk)=0$.  This can be achieved if we take $m=0$
and the coupling matrix $g$ to be antisymmetric, \ie\ $g\ns_{ab}=-g\ns_{ba}$. The system still possesses the
additional parity symmetry of ${\cal P}=\gamma^{45}\, {\cal I}$.

As a specific example, consider the case
\begin{align}
h\ns_{12}=h\ns_{13}=h\ns_{23}&\equiv h\\
h\ns_{14}=h\ns_{24}=h\ns_{34}&\equiv 0\nonumber
\end{align}
and
\begin{align}
g\ns_{12}=g\ns_{13}=g\ns_{23} &\equiv 0 \\
g\ns_{14}=g\ns_{24}=g\ns_{34} &\equiv g\ ,\nonumber
\end{align}
where the elements of $g\ns_{ab}$ below the diagonal follow from antisymmetry.  Then
\begin{align}
\omega(\bfk)&=h\Big[\sin(\theta\ns_1-\theta\ns_2) + \sin(\theta\ns_1-\theta\ns_3) + \sin(\theta\ns_2-\theta\ns_3)\Big]\\
\beta(\bfk)&=g\Big[\sin\theta\ns_1 + \sin\theta\ns_2+ \sin\theta\ns_3\Big]\ .
\end{align}
Note that $\omega(\bfk)$, ${\rm Re}\,\Delta(\bfk)$, and ${\rm Re}\,\beta(\bfk) $ all vanish at the eight
wavevectors $\bfk=\frac{1}{2} (n_1\, \bfb_1+n_2\, \bfb_2 +n_3\, \bfb_3)$ for $n_i=0$ or $1$, which includes
the zone center $\Gamma$, the four $L$-points, and the three $X$-points (which are the three Dirac points).
If $J_{1,2,3,4}=J$, the system remains gapless at three $X$-points because $|{\rm Im}\, \Delta(\bfk)|=0$.
By tuning $J_4\neq J_{1,2,3}=J$, the system can be made gapped, where the
gap at the $X$-points is $|{\rm Im}\> \Delta(\bfk)|=|J_4-J|$.
This situation is the same as the Hamiltonian of the 3D topological
insulator studied by Fu and Kane \cite{fu2007a}.
Following their reasoning, whether or not the insulating states are topological
can be inferred from the parity eigenvalues of $\gamma^{45}$ of the occupied states.
The system is topologically  non-trivial for $J\ns_4>J$, in which case it
exhibit surface modes with an odd number of Dirac cones in open surfaces.

On the other hand, if we relax the requirement of both ${\cal T}$ and 
${\cal P}$ symmetries, and only keep the combined symmetry of ${\cal PT}$.
Then all five coefficients are allowed in eqn. \ref{top4by4}, and each energy level remains doubly degenerate.
However, the analysis in ref. \onlinecite{fu2007a} no longer applies.
We expect then a more diverse set of topological insulators, the study of which will be deferred to a future investigation.
 
\section{Conclusion}
In summary, we have generalized the Kitaev model from the Pauli matrices
to the Clifford algebra of $\Gamma$-matrices.
This enriches the physics of topological states, including the 2D 
chiral spin liquids with non-trivial topological structure, as well as
that of topological spin liquids with time reversal like symmetries.
Possible topological insulating states on the 3D diamond lattice were also discussed.

\begin{acknowledgments}
C. W thank L. Fu, S. Kivelson, X. Qi, S. Ryu, T. Si, X. G. Wen, 
H. Yao, Y. Yu, S. C. Zhang, for helpful discussions on the physics of the Kitaev model and topological
insulators.  C. W. and H. H. H. are supported by the Sloan Research Foundation, 
ARO-W911NF0810291, and NSF-DMR 0804775, and the Academic
Senate research award at UCSD.

{\it Note added.} During the preparation of this work, we learned of 
similar work on a $\Gamma$-matrix extension of the Kitaev model on
the square lattice, in which a gapless algebraic spin liquid
was studied \cite{yao2008}, and also of work on three dimensional
topological phases of the Kitaev model on the diamond lattice \cite{ryu2008}.
\end{acknowledgments}


\begin{thebibliography}{10}
\bibitem{footnote1} Note that while the triangular fluxes combine to yield
$F\nd_{AB'C'D}=-F\nd_{AB'C'}F\nd_{C'DA}$ and $F\nd_{ABC'''D'}=-F\nd_{BC'''D'}F\nd_{D'AB}$, these two minus signs cancel so that the total flux per unit cell
is indeed the product over the six elementary plaquettes.
\end{thebibliography}

\end{document}